# Electrically Tunable Four-Wave-Mixing in Graphene Heterogeneous Fiber for Individual Gas Molecule Detection

Ning An,[#] Teng Tan,[#] Zheng Peng,[#] Chenye Qin, Zhongye Yuan, Lei Bi, Changrui Liao, Yiping Wang, Yunjiang Rao,* Giancarlo Soavi,* and Baicheng Yao*

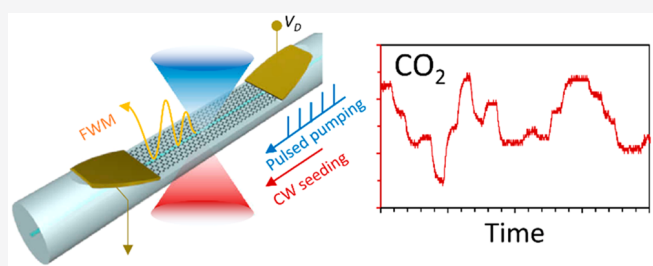

**ABSTRACT:** Detection of individual molecules is the ultimate goal of any chemical sensor. In the case of gas detection, such resolution has been achieved in advanced nanoscale electronic solid-state sensors, but it has not been possible so far in integrated photonic devices, where the weak light−molecule interaction is typically hidden by noise. Here, we demonstrate a scheme to generate ultrasensitive down-conversion four-wave-mixing (FWM) in a graphene bipolar-junction-transistor heterogeneous D-shaped fiber. In the communication band, the FWM conversion efficiency can change steeply when the graphene Fermi level approaches 0.4 eV. In this condition, we exploit our unique two-step optoelectronic heterodyne detection scheme, and we achieve real-time individual gas molecule detection in vacuum. Such combination of graphene strong nonlinearities, electrical tunability, and all-fiber integration paves the way toward the design of versatile high-performance graphene photonic devices.

Graphene displays exceptionally strong light-matter interaction and gate-tunable nonlinear optical properties,[1−3] ranging from difference frequency generation (DFG) based on second order nonlinearities[4,5] to third-order harmonic generation (THG) and four wave mixing (FWM) based on third order nonlinearities, spurring applications such as optical parametric amplifiers, oscillators, switchers, frequency converters, and detectors.[6−10] In particular, refs 8 and 9 have recently demonstrated that the third-order nonlinear susceptibility $\chi^{(3)}$ of graphene can be electrically tuned by more than 1 order of magnitude when moving across THG and FWM multiphoton resonances. Compared to other layered materials, such as transition metal dichalcogenides,[11−13] graphene gapless Dirac Fermions enable a unique broadband response that ranges from terahertz to ultraviolet.[14,15] Moreover, thanks to the atomic flexibility of graphene and related van der Waals integration techniques,[16] graphene can be easily deposited and tailored on-chip or on-fiber, thus further enhancing its potentials for integrated device applications.[17−20] When transferred on waveguides and fibers, graphene gating can also be used to tune the device optical conductivity[21,22] and thus to achieve phase matching. Here, we report electrically tunable frequency down-conversion in graphene heterogeneous silica D-shaped fiber (GhDF), driven by femtosecond degenerate FWM. In this device geometry, the graphene on-fiber is directly exposed to air, thus enabling molecule−graphene interactions and ultrasensitive gas sensing.[23−25] When the graphene $E_F$ is preset at ∼0.4 eV and we use a pump wavelength of ∼1560 nm, FWM in our GhDF device becomes extremely sensitive to molecular adsorption/desorption, enabling label-free individual molecule detection for $NH_3$ and $CO_2$ in vacuum environment.

Figure 1a shows a sketch of our GhDF device. The cladding of a standard single-mode-silica fiber is carefully side-polished, forming a D-shaped evanescent transmission region with nanometer level uniformity and smoothness. A graphene monolayer is grown by using the chemical vapor deposition (CVD) technique and subsequently wet-transferred and patterned on the D-shaped region. Figure 1b shows a microscopy image of our device with a channel width of ∼0.05 mm (a detailed description of the GhDF fabrication and characterization is provided in Supporting Information Note S2). Figure 1c shows the simulated electric field distribution of the fundamental mode inside the GhDF, based on the finite element method. Here the diameter of the fiber core is ∼6 μm. Different works have reported different optimized schemes for light−matter interaction in graphene integrated photonic devices using either TM[26] or TE[27] modes. In this work, we use the TM polarization (electrical field vertical to the

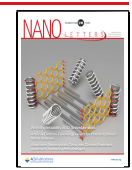

6473

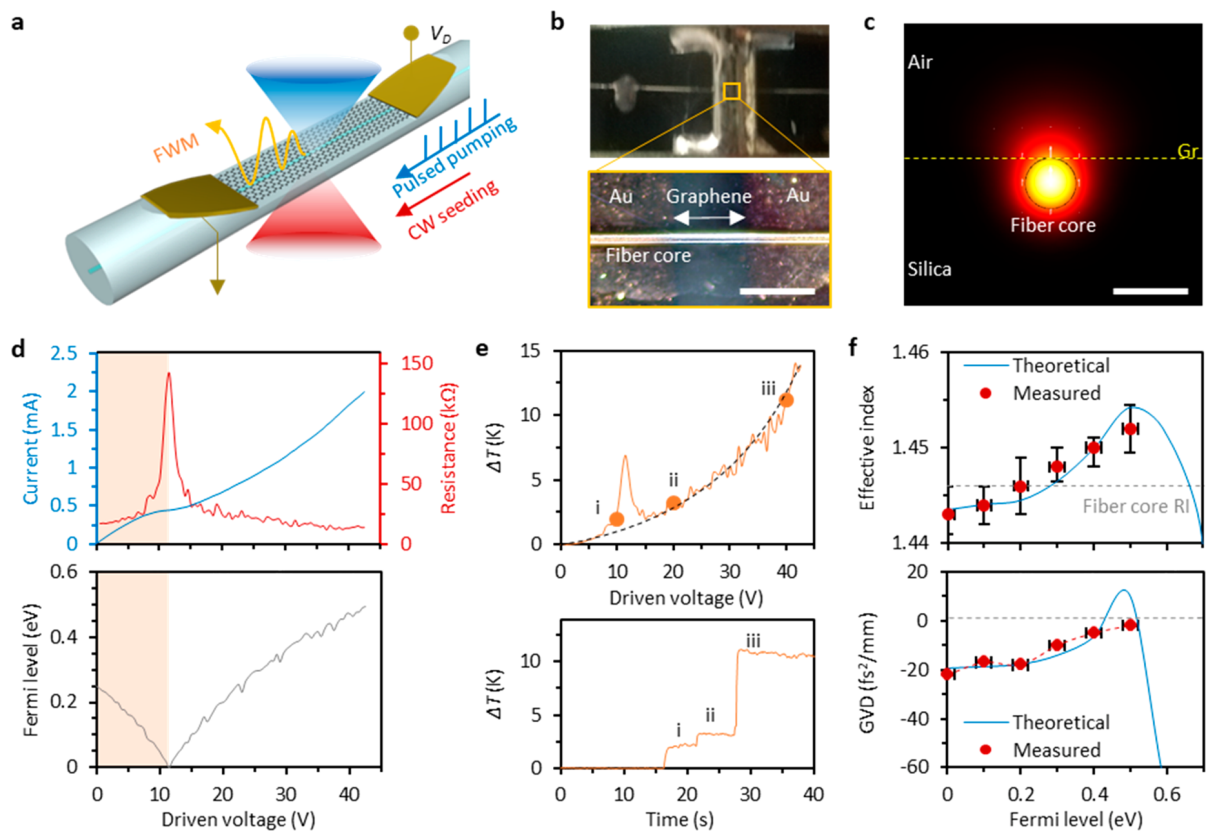

**Figure 1.** Design and implementation of the GhDF for electrically tunable FWM. (a) Sketch of the FWM process. A graphene bipolar junction is fabricated on the side-polished fiber. (b) Top-view optical microscope image of the GhDF device. The graphene channel region is zoomed in, showing the Au−graphene−Au geometry. The bright left-to-right horizontal line is the D-shaped fiber. The scale bar is 50 μm. (c) Sectional view of the electric field distribution in the GhDF with fiber core diameter 6 μm. Scale bar, 10 μm. Here the yellow dashed line highlights the graphene monolayer, and the electric field is vertically polarized (TM polarization). (d) Electronic measurement of the graphene BJT. The kink in the $V_{SD}$–$I_{SD}$ curve at $V_{SD}$ = 11.5 V indicates the position of the Dirac point. (e) Calculated (black dashed curve) and measured (orange curve and dots) increase in temperature when a source-drain voltage is applied to the GhDF. (f) Theoretically modeled refractive index as a function of the Fermi level based on the extracted chromatic dispersion of the device. Measured data points are shown in red dots. The error bar represents the measurement uncertainty estimated from the interferometric measurements under the same conditions.

graphene layer), since this maximizes the light-graphene interaction. In the 1560 nm band, graphene deposited on silica (with $E_F \approx 0.25$ eV) has a refractive index of ~3.34,[22,28] larger than silica (~1.45). A detailed calculation of graphene's refractive index as a function of the Fermi level can be found in refs 22 and 28. In order to achieve electrical tuning of the Fermi level, we deposited two 50/30 nm of Au/Ti electrodes (i.e., source and drain) on the two sides of the graphene monolayer. Figure 1d–f shows the electrical tuning characteristic of our device. By increasing the source-drain voltage ($V_{SD}$), the source-drain current ($I_{SD}$) increases nonlinearly. We notice that in a traditional metal (Ohmic conductor) $I_{SD}$ increases linearly with $V_{SD}$, and thus the resistance ($R = dV_{SD}/dI_{SD}$) is constant. The observed nonlinear increase of $I_{SD}$ in our device is due to the semimetal nature of graphene and the related possibility to electrically tune its Fermi level.[29,30] Figure 1d plots the current ($I_{SD}$) and the channel resistance ($R = dV_{SD}/dI_{SD}$), as a function of $V_{SD}$. Higher current injects more electrons into the graphene monolayer, thus tuning the Fermi level from p-doping to n-doping. The Dirac point in our device corresponds to $V_{SD}$ = 11.5 V, where we observe a maximum channel resistance of $R$ = 142 kΩ. On the other hand, when $V_{SD}$ reaches 40 V, $R$ is smaller than 20 kΩ. Finally, we use the Drude model $\mu = (NRe)^{-1}$ and the quasi Fermi energy equation[31,32] $|E_F| \approx \hbar|v_F|(\pi N)^{-1/2}$ to calculate the $E_F$ in our device as a function of $V_{SD}$. Here $\mu$ indicates the mobility, $N$ is the carrier density in the graphene, $e$ is the electron charge, $\hbar$ is the reduced Planck's constant, $v_F$ is the Fermi velocity, and $R$ is the resistance of the device. The results (Figure 1d) show the $E_F$ tuning from ~−0.25 eV (p-doping) to ~+0.5 eV (n-doping) for $V_{SD}$ tuning between 0 and 45 V. The $E_F$ tuning of our device has also been characterized by using in situ Raman spectroscopy (see Supporting Information Section 2). Typically, the $I_{SD}$ in graphene electrically tunable devices is approximately in the microamperes range and thus it does not induce significant heating. However, the $I_{SD}$ in our device can be about as high as milliamperes, and we thus have to consider the possibility of thermal induced changes of the graphene optical nonlinearities[33] and current induced thermal damage due to Joule heating effect. According to the measured I−V curve, we show the calculated local temperature changes $\Delta T$ versus $V_{SD}$ in Figure 1e. To test our estimate of the Joule heating, we use an infrared thermal imager focusing on the central 1 mm² area to monitor the temperature variations in the GhDF. When $V_{SD}$ increases from 0 to 10, 20, and 40 V, the local temperature increases by 2, 3.2, and 11.2 K, respectively (details are provided in Supporting Information Note S3). Thus, to guarantee stable and reliable results, we limit the maximum $I_{SD}$ to 2 mA, install the GhDF in a vacuum chamber equipped with a probe station, and carefully stabilize the device temperature by using a thermo-electric-cooler (TEC) with 10 mK resolution. Such electrical tunability also influences some of the intrafiber optical parameters: tuning of the graphene conductivity $\sigma_g$ also affects its refractive index $n_g$. By using a stable Mach−Zehnder interferometer, we can measure the electrically tunable effective index and group velocity dispersion of the GhDF.[22] By increasing the graphene Fermi level from 0 to 0.5 eV, the effective index of the GhDF increases from 1.443 to 1.446, meanwhile its anomalous GVD decreases from −20 to −4 fs$^2$/mm at a wavelength of 1560 nm, as shown in Figure 1f.

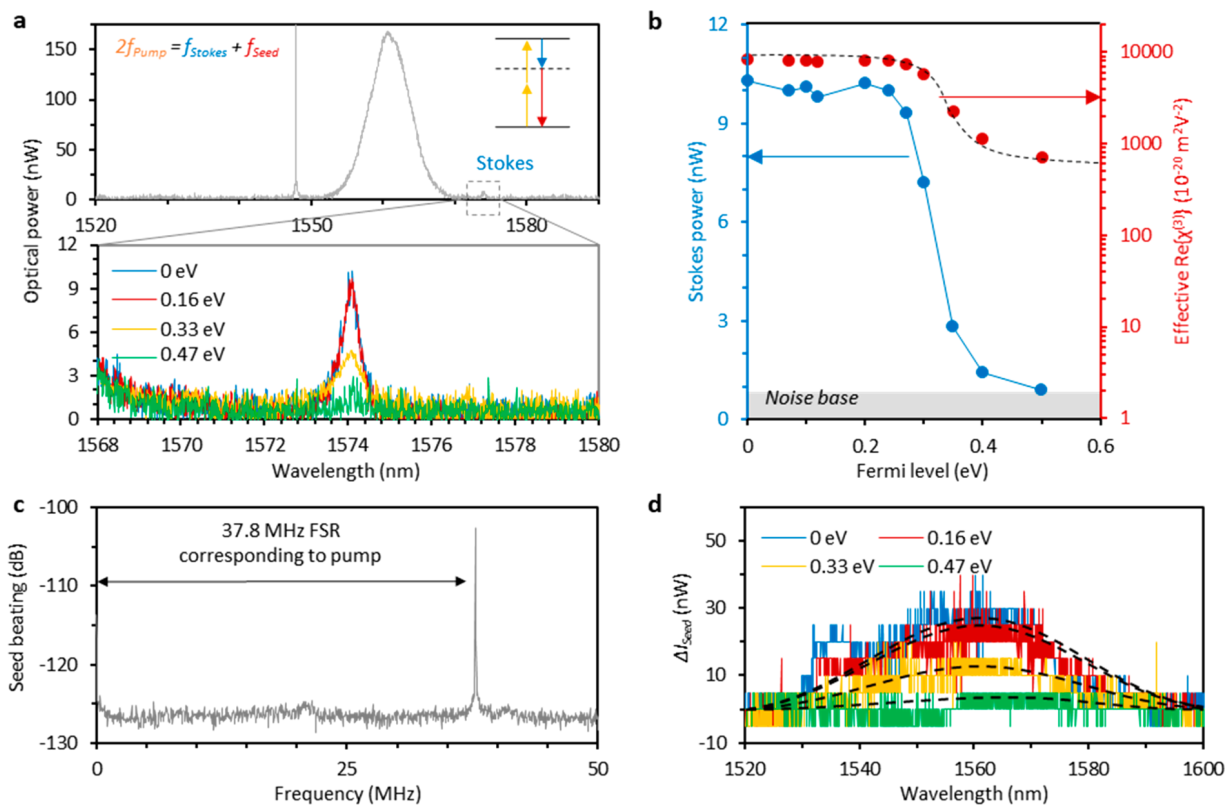

**Figure 2.** Electrically tunable FWM in the graphene heterogeneous D-shaped fiber. (a) Measured spectrum, ranging from 1520 to 1590 nm. Here the central wavelength of the continuous seed, the pulsed pump, and the generated Stokes line are 1548, 1561, and 1573 nm, respectively, allowing energy conservation as $2f_\text{Pump} = f_\text{Seed} + f_\text{Stokes}$. The FWM is electrically tunable when we change the graphene Fermi level via tuning of $V_\text{SD}$. When $E_F$ increases from 0 to 0.47 eV, the Stokes power decreases gradually. (b) Measured Stokes power (blue dots) and the calculated graphene third order nonlinear coefficient $\text{Re}\{\chi^{(3)}\}$ versus the graphene Fermi level. Here the black dashed curve is the moving-average fitting for $\text{Re}\{\chi^{(3)}\}$. (c) Measured beat note of the enhanced seed at 37.8 MHz. (d) The seed enhancement varies when tuning the $\lambda_\text{Seed}$ from 1500 to 1600 nm. When the detuning approaches 0, the FWM has the highest conversion efficiency.

Figure 2 shows the FWM generation and electrical tuning of our device. FWM is obtained with a mode-locked femtosecond laser (central wavelength 1561 nm, or central frequency $f_\text{Pump}$ = 192.19 THz) as the optical pump and a tunable continuous-wave (tunable range 1480−1610 nm, or $f_\text{Seed}$ = 202.7−186.34 THz) as the optical seed. Once the phase-matching condition is satisfied, FWM will generate new photons (Stokes line) at frequency $2f_\text{Pump} - f_\text{Seed} < f_\text{Pump}$, as shown in Figure 2a. The power of the generated Stokes beam is defined by the third nonlinear coefficient $\chi^{(3)}$, which in turn is a function of the graphene $E_F$. When fixing the pump-seed detuning to 13 nm ($\lambda_\text{Pump}$ = 1561 nm, $\lambda_\text{Seed}$ = 1548 nm) and increasing the graphene $E_F$ from 0 to 0.47 eV via $V_\text{SD}$, we observe a decrease of the Stokes power ($\lambda_\text{Stokes}$ = 1574 nm) from ∼10 nW to <2 nW (down to noise base), as shown in Figure 2b. For further measurements on the FWM gate tuneability, see Supporting Information Movie 1. The stabilized mode locked fiber laser launches a maximum average power of ∼3 mW into the GhDF, corresponding to a maximum peak power of ∼0.3 kW (37.8 MHz repetition, 265 fs pulse duration). Thus, for the 13 nm pump-seed detuning, the nonlinear conversion efficiency is ∼−55 dB when $E_F$ < 0.2 eV, in agreement with recent results on FWM from graphene on waveguides.[19,20] We notice that the observed modulation of the Stokes power with the external voltage (Figure 2b) is due to both multiphoton resonant transitions and tuning of the phase matching condition (a higher $E_F$ induces a larger $n_g$, resulting in a smaller nonlinear gain). Note that the multiphoton resonance does not occur exactly at $\hbar\omega = 2E_F$, being $\hbar\omega \approx 0.8$ eV for our pump wavelength (1560 nm). This is due to the effect of the high electronic temperature reached in nonlinear experiments and quantum interference between the multiphoton resonant transitions contributing to the FWM nonlinear optical susceptibility, as also observed in previous studies.[9,20] Finally, we plot the calculated graphene $\text{Re}\{\chi^{(3)}\}$ (red dots in Figure 2b). When $E_F$ increases from 0 to 0.5 eV, the $\text{Re}\{\chi^{(3)}\}$ of the GhDF is tuned from $8 \times 10^{-17}$ to $6 \times 10^{-18}$ m$^2$/V$^2$, considering the transverse mode area in the fiber of ∼40 $\mu$m$^2$. Energy conversion in an FWM process can be tested either by measuring the pump depletion ($\Delta P_\text{Pump}$) or the seed enhancement ($\Delta P_\text{Seed}$), given the relation $2\Delta P_\text{Pump} = \Delta P_\text{Seed} + P_\text{Stokes}$. We measure the self-beat of the FWM enhanced seed by using a balanced photodetector (Figure 2c). Without FWM, the continuous wave seeding cannot have any beat note. On the other hand, the FWM process transfers energy at a fixed repetition rate of 37.8 MHz from the pulsed pump to the CW seed. We thus observe oscillations (at 37.8 MHz rate) in the seed enhancement on top of its DC average power, as shown also in Supporting Information Figures S7 and S8. Measurements of $\Delta P_{Seed}$ also enable us to identify the nonlinear conversion efficiency as a function of the pump-seed detuning. In order to test this, we fix the pump wavelength and scan the seed wavelength from 1500 to 1600 nm while monitoring $\Delta P_{Seed}$ (Figure 2d). As expected from the phase matching condition, larger detuning gives smaller nonlinear enhancement. When $|\lambda_{Pump} - \lambda_{Seed}|$ approaches 0, we observe the highest FWM conversion efficiency. At this point, we estimate $\gamma Re\{\chi^{(3)}\}$ of ∼$10^{-15}$ m$^2$ V$^{-2}$, in agreement with previous studies.[9,20]

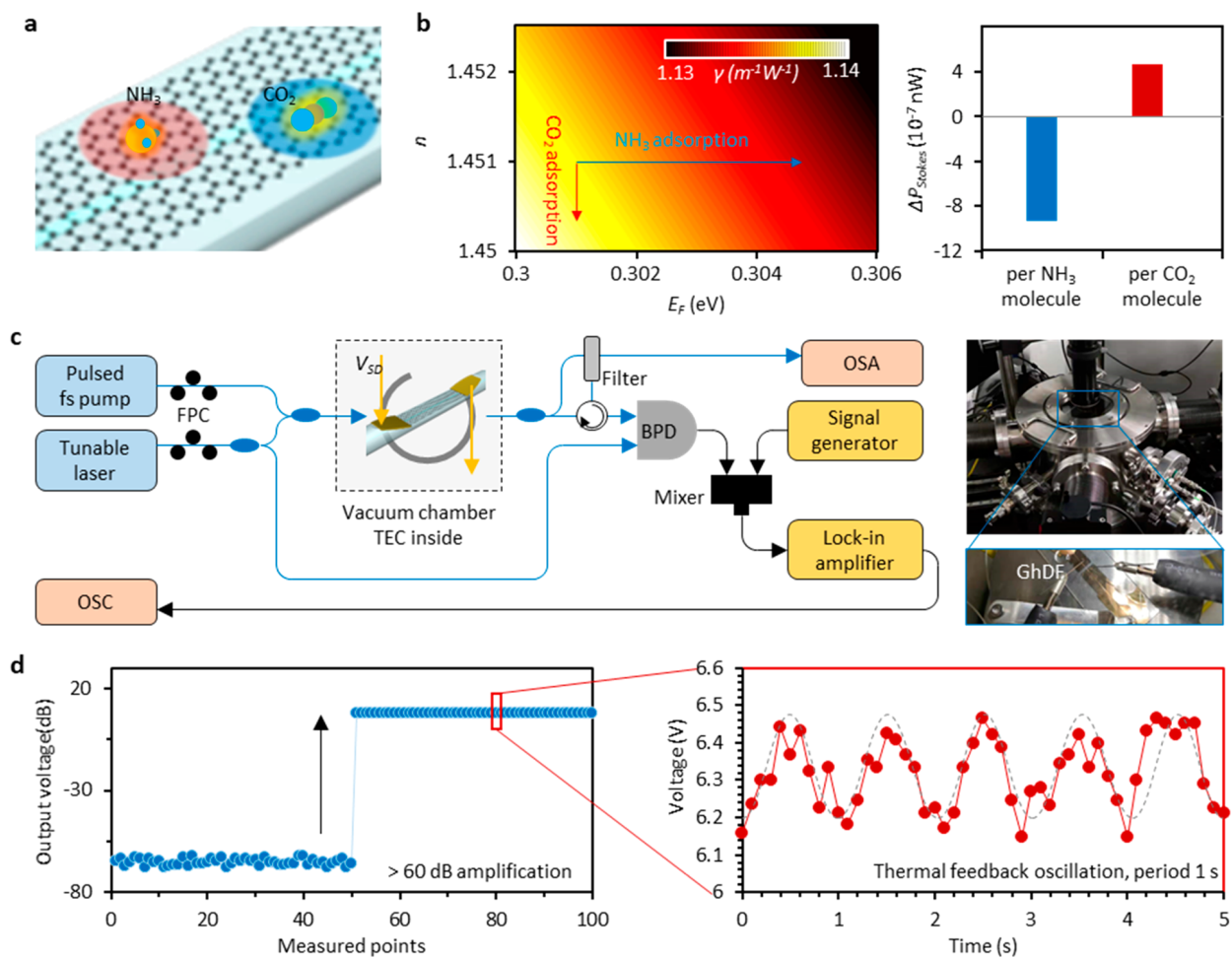

**Figure 3.** Mechanism and experimental setup for gas sensing. (a) Sketch of $NH_3/CO_2$ adsorption on graphene. The gas molecule adsorption induces either charge transfer or structural impurity. (b) Simulation of the three-dimensional correlation: the Fermi level ($E_F$), the effective refractive index of the GhDF ($n$), and the third order nonlinear coefficient ($\gamma$). $NH_3$ adsorption mainly changes the $E_F$, while $CO_2$ adsorption mainly changes the $n$. Each $NH_3/CO_2$ molecular adsorption induces $\Delta P_{Stokes}$ of $\sim -9/+4.4 \times 10^{-7}$ nW. (c) Experimental setup. The GhDF is fixed in a temperature-controlled vacuum chamber. Here, BPD, balanced photodetector; OSA, optical spectrum analyzer; OSC, oscilloscope. (d) Performance of the lock-in amplification with amplification of the modulated signal >60 dB. (Inset) The thermal feedback induced uncertainty is ±0.15 mV with thermal noise frequency ∼1 Hz.

The steep change in the FWM efficiency when the graphene $E_F$ approaches 0.4 eV can be used as an ultrasensitive tool for chemical sensing. Indeed, gas adsorption on graphene can modulate the $E_F$ via both charge transfer and impurity doping[23,24,34] and, as a consequence, it will affect $\Delta P_{Seed}$ during the FWM experiment. Figure 3a schematically shows the case of $NH_3$ and $CO_2$ molecules attached to the GhDF. In the limiting case of individual molecular interaction, $NH_3$ will act as a donor material with two electrons per molecule, thus directly affecting the graphene's $E_F$.[23,35] Moreover, we highlight that since the sensing mechanism for $NH_3$ molecule is based on the contribution of lone electron pairs from the N atom in the $NH_3$ molecule, our sensing scheme could be used also for the detection of other amino molecules (e.g., $NH_2CH_3$ and $N(CH_2CH_3)_3$). On the other hand, $CO_2$ adsorbed on the graphene introduces scattering impurities, thus changing the refractive index of graphene.[24] Figure 3b shows the quantitative simulations of the nonlinear gain coefficient $\gamma$ of the GhDF when changing the graphene's $E_F$ and the GhDF's refractive index $n$, being $\gamma = n_2\omega_1/(cA_{eff})$ and $n_2 = 3\pi f Re\{\chi^{(3)}\}/4\varepsilon_0 n^2 c^2 A_{eff}$ the nonlinear refractive index. In this equation, $n = 1.46$ is the (linear) refractive index of the GhDF (without adsorbed molecules), $f = 192.3$ THz is the optical frequency, $c$ is the speed of light in vacuum, $\varepsilon_0$ is the vacuum permittivity, and $A_{eff} \approx 40$ $\mu m^2$ is the effective transverse mode area of the GhDF. We thus obtain $\gamma \approx 1.135$ m$^{-1}$ W$^{-1}$. In this framework, $NH_3$ adsorption increases the graphene Fermi level (and thus affects mainly $Re\{\chi^{(3)}\}$) while $CO_2$ adsorption decreases the (linear) effective index ($n$) of the GhDF. Supposing that graphene is predoped around 0.4 eV, and the effective area of the device is $\sim 5 \times 10^{-10}$ m$^2$, then (i) single $NH_3$ molecule adsorption will decrease the carrier density by $4 \times 10^9$ m$^{-2}$, leading to an $E_F$ of $\sim 9.8 \times 10^{-9}$ eV per molecule. Here, we have used $|E_F| \approx \hbar |v_F| (\pi N)^{-1/2}$, where $v_F = 10^6$ m/s is the Fermi velocity. Thus, based on the results shown in Figure 2b ($E_F$ versus $P_{Stokes}$ relation $\approx -90$ nW/eV), in our device a single $NH_3$ molecule adsorption event will induce a reduction of the Stokes power of $\sim 9 \times 10^{-7}$ nW per molecule. (ii) Individual $CO_2$ molecule adsorption will increase the graphene resistance via impurity doping,[24] thus reducing the effective refractive index by $10^{-12}$, estimated by using FET calculations. This leads to an increase of $\gamma$ of $\sim 2.4 \times 10^{-7}$ m$^{-1}$ W$^{-1}$. Thus, individual $CO_2$ molecule adsorption will increase the Stokes power by $\sim 4.4 \times 10^{-7}$ nW per molecule (see also Supporting Information Note S1). Since the expected changes in the Stokes power upon single molecule adsorption are considerably lower compared to our noise level of $\sim 0.1$ nW, in order to detect individual adsorption/desorption dynamics we performed heterodyne optoelectronic measurements in vacuum, as depicted in Figure 3c. Before launching into the GhDF, we use a 1:1 fiber coupler to mix the femtosecond pulsed pump and the continuous wave seed. The polarization of both is optimized independently. The GhDF is fixed in a vacuum chamber equipped with a 4-probe station, which is used for electrical tuning of the GhDF. The volume of the vacuum chamber is 8 L with maximum vacuity of $10^{-5}$ Torr. Inside the chamber, we use a thermoelectric cooler (TEC) to control the temperature with 10 mK accuracy. Since our single molecule detection is based on the precise measurement of $\Delta P_{Seed}$ (i.e., we need to measure $\Delta P_{Seed} \approx 10^{-7}$ nW), in our experiments we use a fiber Bragg grating based tunable filter (bandwidth 0.1 nm centered at 1547.84 nm to filter the seed wavelength) and balanced detection to remove the DC components. Afterward, we use a stable RF generator to beat the repetition frequency of $\Delta P_{Seed}$, for further amplification and frequency down conversion. The two-step beating-based heterodyne amplification finally boosts the minor signal alteration over >60 dB (Figure 3d). With such amplification (>6 orders of magnitude), we are able to detect $\Delta P_{Seed} \approx 10^{-7}$ nW. In other words, the heterodyne-based lock-in amplification technique is a powerful tool for weak signal extraction, which is uniquely suitable for nonlinear detection and high precision sensing.[5,36,37] We note that such measurement is extremely sensitive to power fluctuations, thus we have to carefully stabilize the optical sources and the device temperature. In this regard, we apply a thermal feedback with response rate of 1 Hz. However, despite the applied stabilization we can still observe thermally induced fluctuations of ±0.15 V. Hence, in order to avoid cross-influence during gas sensing experiment, we carefully control the gas concentration with nM/L precision, making sure that the gas molecular on/off dynamics has a rate >100 Hz, much faster than the thermal fluctuation. Gas preparation is discussed in Supporting Information Note 2, and Supporting Information Figure S8 plots the spectral characterization and noise suppression of this optoelectronic heterodyne system.

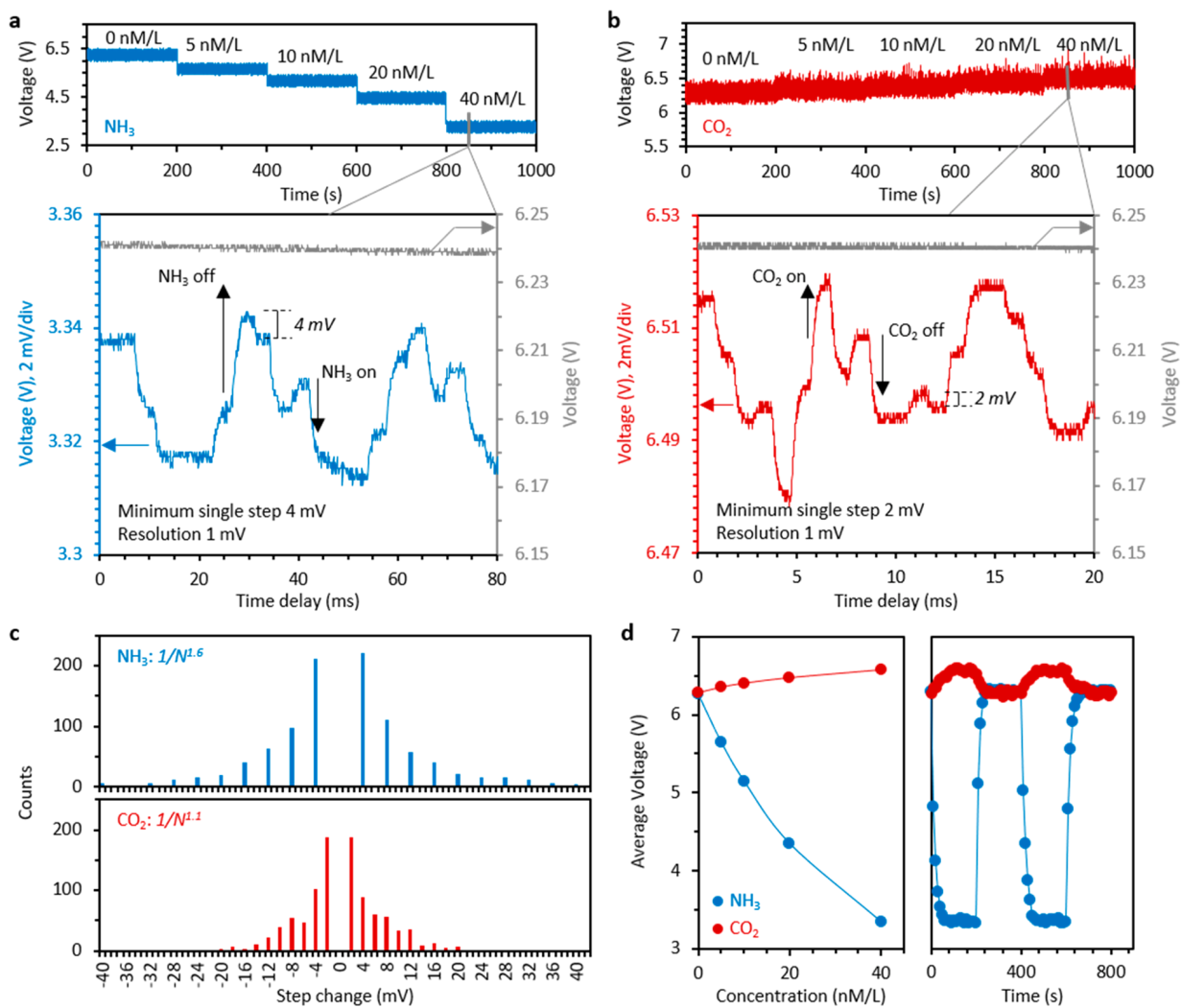

**Figure 4.** Gas sensing performance. (a,b) Locked-in amplified voltage $\Delta P_{Seed}$, when sensing $NH_3$ (blue curve) and $CO_2$ (red curve). The $NH_3$ adsorption decreases the FWM efficiency while $CO_2$ adsorption increases the FWM efficiency. When zooming-in the curves over tens of milliseconds, molecular on/off dynamics become evident. Here the gray curve shows the reference voltage when there is no gas molecule in the vacuum chamber (0 nM/L). (c) In the dynamically stable region, we count the molecular on/off (adsorption/desorption) events. The obtained statistics follow a power law, as discussed in the main text. (d,e) (Left) Gas concentration versus average measured voltage. The maximum macrosensitivity for $NH_3$ and $CO_2$ are ~120 and 15 mV/nM. (Right) By electrically tuning the graphene temperature, our device shows good recoverability.

Figure 4 shows the sensing capability of our GhDF device. Figure 4a shows the changes in the locked-in amplified signal $\Delta P_{Seed}$ upon an increment of the $NH_3$ gas concentration from 0 to 40 nM/L, while Figure 4b shows the case of $CO_2$ gas injection. As explained in Figure 3, the $NH_3$ ($CO_2$) adsorption leads to a lower (higher) FWM efficiency. For an integration time of 10 μs, the maximum sensitivity of our GhDF is higher than 120 mV/nM for $NH_3$ and 15 mV/nM for $CO_2$ (left panel in Figure 4d). The uncertainty of our lock-in amplification based measurement is ±1 mV. In this vacuum environment, 1 nM means 0.6 molecules/$\mu m^3$, and thus, considering the effective graphene–gas interaction area of our device of 500 $\mu m^2$, there could be tens to hundreds of molecules adsorbed on the graphene surface simultaneously. Thus, in order to characterize the molecular dynamics, we zoom-in the time trace to detect individual molecule adsorption events (Figure 4a,b). Individual gas molecules attaching or detaching from the graphene will induce discrete steplike changes with integer common denominator on a < 10 ms time-scale (i.e., the inverse of our gas on/off rate >100 Hz). Experimental results (Figure 4a,b) show that in the case of $NH_3$ the smallest voltage step is 4 mV, while for $CO_2$ the smallest step is 2 mV. For the detection of the Stokes Power, we used an InGaAs balanced photodetector (Newport Nirvana 2017), which has a conversion gain of $\sim 10^6$ V/W at 1550 nm. Thus, considering the amplification due to lock-in detection of 66.2 dB (Figure 3), the measured electrical signal will be ~2 mV for individual $CO_2$ molecules (4.4 × 10−7 nW) and ~4 mV for individual $NH_3$ molecules (9 × 10−7 nW), in agreement with experimental results. Such discrete steps strongly point toward individual molecule adsorption/desorption events and demonstrate the individual molecule detection capability of our system. The $NH_3$ on/off interval (~5 ms per step) is significantly larger than the $CO_2$ on/off interval (~1 ms per step), because the $NH_3$–graphene bonding is tighter compared to $CO_2$–graphene, that is, the $NH_3$ adsorption on graphene is more stable (see also Supporting Information Figure S9). Moreover, we counted the molecular on/off events (total number 1000, Figure 4c). We estimated the number of molecules that are adsorbed/desorbed on our device as the ratio between the observed steps (in mV) and the minimum step expected for individual molecule adsorption/desorption events, that is, 4 and 2 mV for $NH_3$ and $CO_2$ respectively.

When the GhDF is exposed either to $NH_3$ or to $CO_2$ in the quasi-static state, large steps were rare (such as 10+ molecules on/off), whereas unit steps were dominant. These statistical results obey a power-law distribution, which is also a sign of individual molecule adsorption events.[23] In particular, we can fit the molecules-counts correlation of $NH_3$ and $CO_2$ with the power-law $1/N^{1.6}$ and $1/N^{1.1}$, respectively. This result further suggests that, as previously discussed, adsorption on graphene is stronger in the case of $NH_3$ compared to $CO_2$. Additional measurements down to 10 pM/L concentration are shown in Supporting Information Figure S10. Finally, we note that adsorbed molecules can be efficiently released from the graphene surface by temperature-induced annealing. As shown in the right panel of Figure 4d, by tuning the device temperature between 300 and 370 K we achieve almost 100% recovery of the initial graphene properties for both $NH_3$ and $CO_2$. The recovery time constant is less than 1 min.

In summary, electrically tunable FWM was achieved in a graphene heterogeneous D-shaped fiber device. By changing the graphene Fermi level from 0 to 0.5 eV, the $Re\{\chi^{(3)}\}$ of graphene can be modulated over 1 order of magnitude from 8 × $10^{-17}$ to 8 × $10^{-18}$ $m^2$ $V^{-2}$. A steep change (increase/decrease) of the FWM efficiency when the Fermi level approaches 0.4 eV allows for ultrasensitive characterization of third order nonlinearity of graphene. We exploit this feature to develop a fully integrated fiber-optic gas sensor device with gas molecule detection capability. As an application, the nonlinearity enhanced GhDF shows high performance as a label free and cheap fiber-optic sensor at room temperature with individual molecule sensitivity. Moreover, our integrated device, which combines graphene optical nonlinearity, electrical tunability, and on-fiber integration, will pave the way for other integrated photonic applications ranging from parametric amplification, photonic switching, frequency modulation, and high precision measurement.

## ■ ASSOCIATED CONTENT

### ⓈSupporting Information

The Supporting Information is available free of charge at https://pubs.acs.org/doi/10.1021/acs.nanolett.0c02174.

> (S1) Electrically tunable four-wave-mixing in the graphene based D-shaped fiber; (S2) fabrication and characterization of the GhDF and device/setup preparation for sensing; (S3) measurements calibration and low concentration measurements; additional references and figures (PDF)
>
> FWM gate tunability (MP4)

## ■ AUTHOR INFORMATION

### Corresponding Authors

Yunjiang Rao − *Key Laboratory of Optical Fiber Sensing and Communications (Education Ministry of China), University of Electronic Science and Technology of China, Chengdu 610054, China; Research Centre of Optical Fiber Sensing, Zhejiang Laboratory, Hangzhou 310000, China*; Email: yjrao@uestc.edu.cn

Giancarlo Soavi − *Institute of Solid State Physics, Abbe Center of Photonics, Friedrich-Schiller-University Jena, Jena 07743, Germany*; Email: giancarlo.soavi@uni-jena.de

Baicheng Yao − *Key Laboratory of Optical Fiber Sensing and Communications (Education Ministry of China), University of Electronic Science and Technology of China, Chengdu 610054,*

China; orcid.org/0000-0001-8368-5815;
Email: yaobaicheng@uestc.edu.cn

**Authors**

Ning An − *Key Laboratory of Optical Fiber Sensing and Communications (Education Ministry of China), University of Electronic Science and Technology of China, Chengdu 610054, China*

Teng Tan − *Key Laboratory of Optical Fiber Sensing and Communications (Education Ministry of China), University of Electronic Science and Technology of China, Chengdu 610054, China; Research Centre of Optical Fiber Sensing, Zhejiang Laboratory, Hangzhou 310000, China*

Zheng Peng − *State Key Laboratory of Electronic Thin Film and Integrated Devices, University of Electronic Science and Technology of China, Chengdu 610054, China*

Chenye Qin − *Key Laboratory of Optical Fiber Sensing and Communications (Education Ministry of China), University of Electronic Science and Technology of China, Chengdu 610054, China*

Zhongye Yuan − *Key Laboratory of Optical Fiber Sensing and Communications (Education Ministry of China), University of Electronic Science and Technology of China, Chengdu 610054, China*

Lei Bi − *State Key Laboratory of Electronic Thin Film and Integrated Devices, University of Electronic Science and Technology of China, Chengdu 610054, China;* orcid.org/0000-0002-2698-2829

Changrui Liao − *Guangdong and Hong Kong Joint Research Center for Optical Fiber Sensors, Shenzhen University, Shenzhen 518060, China;* orcid.org/0000-0003-3669-5054

Yiping Wang − *Guangdong and Hong Kong Joint Research Center for Optical Fiber Sensors, Shenzhen University, Shenzhen 518060, China*

Complete contact information is available at:
https://pubs.acs.org/10.1021/acs.nanolett.0c02174

**Author Contributions**

B.Y. led and Y.R. supervised this work. B.Y. and G.S. led the in-principle analysis, calculations, and simulations. N.A., T.T., C.Q., and B.Y. conducted the optoelectronic measurements. N.A., C.L. and Y.W. fabricated and tested the D-shaped fiber samples. N.A., Z.P., C.Q., and L.B. performed the graphene heterogeneous device fabrication and characterization. T.T. and Z.Y. contributed the femtosecond pump locking and stabilization. N.A., B.Y., and T.T. built the experimental setup for sensing. B.Y., G.S., N.A., T.T., L.B., C.W.W., and Y.R. performed the measured data analysis. All authors discussed the results. B.Y., G.S., N.A., and Y.R. prepared the manuscript.

**Author Contributions**

#N.A., T.T., and Z.P. contributed equally.

**Notes**

The authors declare no competing financial interest.

## ACKNOWLEDGMENTS

We thank Dr. Handing Xia for the preparation of femtosecond laser pump. This work at University of Electronic Science and Technology of China is supported by the National Natural Science Foundation of China (Grants 61975025, 61705032, and 51972044), Sichuan Provincial Science and Technology Department (Grant 2019YFH0154), and Ministry of Science and Technology of the People's Republic of China (MOST) (Grants 2016YFA0300802 and 2018YFE0109200). This project has received funding from the European Union's Horizon 2020 research and innovation program under Grant Agreement GrapheneCore3 881603.

## REFERENCES

(1) Lim, G.-K.; Chen, Z.-L.; Clark, J.; Goh, R. G. S.; Ng, W.-H.; Tan, H.-W.; Friend, R. H.; Ho, P. K. H.; Chua, L.-L. Giant Broadband Nonlinear Optical Absorption Response in Dispersed Graphene Single Sheets. *Nat. Photonics* **2011**, *5* (9), 554−560.

(2) Bonaccorso, F.; Sun, Z.; Hasan, T.; Ferrari, A. C. Graphene Photonics and Optoelectronics. *Nat. Photonics* **2010**, *4* (9), 611−622.

(3) Koppens, F. H. L.; Chang, D. E.; García De Abajo, F. J. Graphene Plasmonics: A Platform for Strong Light-Matter Interactions. *Nano Lett.* **2011**, *11* (8), 3370−3377.

(4) Constant, T. J.; Hornett, S. M.; Chang, D. E.; Hendry, E. All-Optical Generation of Surface Plasmons in Graphene. *Nat. Phys.* **2016**, *12* (2), 124−127.

(5) Yao, B.; Liu, Y.; Huang, S.-W.; Choi, C.; Xie, Z.; Flor Flores, J.; Wu, Y.; Yu, M.; Kwong, D.-L.; Huang, Y.; Rao, Y.; Duan, X.; Wong, C. W. Broadband Gate-Tunable Terahertz Plasmons in Graphene Heterostructures. *Nat. Photonics* **2018**, *12* (1), 22−28.

(6) Hendry, E.; Hale, P. J.; Moger, J.; Savchenko, A. K.; Mikhailov, S. A. Coherent Nonlinear Optical Response of Graphene. *Phys. Rev. Lett.* **2010**, *105* (9), 097401.

(7) Hong, S.-Y.; Dadap, J. I.; Petrone, N.; Yeh, P.-C.; Hone, J.; Osgood, R. M. Optical Third-Harmonic Generation in Graphene. *Phys. Rev. X* **2013**, *3* (2), 021014.

(8) Soavi, G.; Wang, G.; Rostami, H.; Purdie, D. G.; De Fazio, D.; Ma, T.; Luo, B.; Wang, J.; Ott, A. K.; Yoon, D.; Bourelle, S. A.; Muench, J. E.; Goykhman, I.; Dal Conte, S.; Celebrano, M.; Tomadin, A.; Polini, M.; Cerullo, G.; Ferrari, A. C. Broadband, Electrically Tunable Third-Harmonic Generation in Graphene. *Nat. Nanotechnol.* **2018**, *13* (7), 583−588.

(9) Jiang, T.; Huang, D.; Cheng, J.; Fan, X.; Zhang, Z.; Shan, Y.; Yi, Y.; Dai, Y.; Shi, L.; Liu, K.; Zeng, C.; Zi, J.; Sipe, J. E.; Shen, Y. R.; Liu, W. T.; Wu, S. Gate-Tunable Third-Order Nonlinear Optical Response of Massless Dirac Fermions in Graphene. *Nat. Photonics* **2018**, *12* (7), 430−436.

(10) Tan, T.; Jiang, X.; Wang, C.; Yao, B.; Zhang, H. 2D Material Optoelectronics for Information Functional Device Applications: Status and Challenges. *Adv. Sci.* **2020**, *7* (11), 2000058.

(11) Mak, K. F.; Shan, J. Photonics and Optoelectronics of 2D Semiconductor Transition Metal Dichalcogenides. *Nat. Photonics* **2016**, *10*, 216−226.

(12) Koppens, F. H. L.; Mueller, T.; Avouris, P.; Ferrari, A. C.; Vitiello, M. S.; Polini, M. Photodetectors Based on Graphene, Other Two-Dimensional Materials and Hybrid Systems. *Nat. Nanotechnol.* **2014**, *9*, 780−793.

(13) Wu, S.; Buckley, S.; Schaibley, J. R.; Feng, L.; Yan, J.; Mandrus, D. G.; Hatami, F.; Yao, W.; Vučković, J.; Majumdar, A.; Xu, X. Monolayer Semiconductor Nanocavity Lasers with Ultralow Thresholds. *Nature* **2015**, *520* (7545), 69−72.

(14) Nair, R. R.; Blake, P.; Grigorenko, A. N.; Novoselov, K. S.; Booth, T. J.; Stauber, T.; Peres, N. M. R.; Geim, A. K. Fine Structure Constant Defines Visual Transparency of Graphene. *Science* **2008**, *320* (5881), 1308.

(15) Mak, K. F.; Sfeir, M. Y.; Wu, Y.; Lui, C. H.; Misewich, J. A.; Heinz, T. F. Measurement of the Optical Conductivity of Graphene. *Phys. Rev. Lett.* **2008**, *101* (19), 196405.

(16) Liu, Y.; Huang, Y.; Duan, X. Van Der Waals Integration before and beyond Two-Dimensional Materials. *Nature* **2019**, *567*, 323−333.

(17) Martinez, A.; Sun, Z. Nanotube and Graphene Saturable Absorbers for Fibre Lasers. *Nat. Photonics* **2013**, *7*, 842−845.

(18) Sun, Z.; Martinez, A.; Wang, F. Optical Modulators with 2D Layered Materials. *Nat. Photonics* **2016**, *10*, 227−238.

(19) Gu, T.; Petrone, N.; McMillan, J. F.; Van Der Zande, A.; Yu, M.; Lo, G. Q.; Kwong, D. L.; Hone, J.; Wong, C. W. Regenerative

Oscillation and Four-Wave Mixing in Graphene Optoelectronics. *Nat. Photonics* **2012**, *6* (8), 554−559.

(20) Alexander, K.; Savostianova, N. A.; Mikhailov, S. A.; Kuyken, B.; Van Thourhout, D. Electrically Tunable Optical Nonlinearities in Graphene-Covered SiN Waveguides Characterized by Four-Wave Mixing. *ACS Photonics* **2017**, *4* (12), 3039−3044.

(21) Phare, C. T.; Daniel Lee, Y. H.; Cardenas, J.; Lipson, M. Graphene Electro-Optic Modulator with 30 GHz Bandwidth. *Nat. Photonics* **2015**, *9* (8), 511−514.

(22) Yao, B.; Huang, S. W.; Liu, Y.; Vinod, A. K.; Choi, C.; Hoff, M.; Li, Y.; Yu, M.; Feng, Z.; Kwong, D. L.; Huang, Y.; Rao, Y.; Duan, X.; Wong, C. W. Gate-Tunable Frequency Combs in Graphene-Nitride Microresonators. *Nature* **2018**, *558* (7710), 410−414.

(23) Schedin, F.; Geim, A. K.; Morozov, S. V.; Hill, E. W.; Blake, P.; Katsnelson, M. I.; Novoselov, K. S. Detection of Individual Gas Molecules Adsorbed on Graphene. *Nat. Mater.* **2007**, *6* (9), 652−655.

(24) Sun, J.; Muruganathan, M.; Mizuta, H. Room Temperature Detection of Individual Molecular Physisorption Using Suspended Bilayer Graphene. *Sci. Adv.* **2016**, *2* (4), No. e1501518.

(25) Yao, B.; Yu, C.; Wu, Y.; Huang, S.-W.; Wu, H.; Gong, Y.; Chen, Y.; Li, Y.; Wong, C. W.; Fan, X.; Rao, Y. Graphene-Enhanced Brillouin Optomechanical Microresonator for Ultrasensitive Gas Detection. *Nano Lett.* **2017**, *17* (8), 4996−5002.

(26) Bao, Q.; Zhang, H.; Wang, B.; Ni, Z.; Lim, C. H. Y. X.; Wang, Y.; Tang, D. Y.; Loh, K. P. Broadband Graphene Polarizer. *Nat. Photonics* **2011**, *5* (7), 411−415.

(27) Zapata, J. D.; Steinberg, D.; Saito, L. A. M.; de Oliveira, R. E. P.; Cárdenas, A. M.; de Souza, E. A. T. Efficient Graphene Saturable Absorbers on D-Shaped Optical Fiber for Ultrashort Pulse Generation. *Sci. Rep.* **2016**, *6* (1), 20644.

(28) Yao, B.; Wu, Y.; Wang, Z.; Cheng, Y.; Rao, Y.; Gong, Y.; Chen, Y.; Li, Y. Demonstration of Complex Refractive Index of Graphene Waveguide by Microfiber-Based Mach−Zehnder Interferometer. *Opt. Express* **2013**, *21* (24), 29818.

(29) Meric, I.; Han, M. Y.; Young, A. F.; Ozyilmaz, B.; Kim, P.; Shepard, K. L. Current Saturation in Zero-Bandgap, Top-Gated Graphene Field-Effect Transistors. *Nat. Nanotechnol.* **2008**, *3* (11), 654−659.

(30) Schwierz, F. Graphene Transistors. *Nat. Nanotechnol.* **2010**, *5* (7), 487−496.

(31) Das, A.; Pisana, S.; Chakraborty, B.; Piscanec, S.; Saha, S. K.; Waghmare, U. V.; Novoselov, K. S.; Krishnamurthy, H. R.; Geim, A. K.; Ferrari, A. C.; Sood, A. K. Monitoring Dopants by Raman Scattering in an Electrochemically Top-Gated Graphene Transistor. *Nat. Nanotechnol.* **2008**, *3* (4), 210−215.

(32) Tan, Y.-W.; Zhang, Y.; Bolotin, K.; Zhao, Y.; Adam, S.; Hwang, E. H.; Das Sarma, S.; Stormer, H. L.; Kim, P. Measurement of Scattering Rate and Minimum Conductivity in Graphene. *Phys. Rev. Lett.* **2007**, *99* (24), 246803.

(33) Soavi, G.; Wang, G.; Rostami, H.; Tomadin, A.; Balci, O.; Paradisanos, I.; Pogna, E. A. A.; Cerullo, G.; Lidorikis, E.; Polini, M.; Ferrari, A. C. Hot Electrons Modulation of Third-Harmonic Generation in Graphene. *ACS Photonics* **2019**, *6* (11), 2841−2849.

(34) Chen, J. H.; Jang, C.; Adam, S.; Fuhrer, M. S.; Williams, E. D.; Ishigami, M. Charged-Impurity Scattering in Graphene. *Nat. Phys.* **2008**, *4* (5), 377−381.

(35) Saffarzadeh, A. Modeling of Gas Adsorption on Graphene Nanoribbons. *J. Appl. Phys.* **2010**, *107* (11), 114309.

(36) Mauranyapin, N. P.; Madsen, L. S.; Taylor, M. A.; Waleed, M.; Bowen, W. P. Evanescent Single-Molecule Biosensing with Quantum-Limited Precision. *Nat. Photonics* **2017**, *11* (8), 477−481.

(37) Cao, Z.; Yao, B.; Qin, C.; Yang, R.; Guo, Y.; Zhang, Y.; Wu, Y.; Bi, L.; Chen, Y.; Xie, Z.; Peng, G.; Huang, S.-W.; Wong, C. W.; Rao, Y. Biochemical Sensing in Graphene-Enhanced Microfiber Resonators with Individual Molecule Sensitivity and Selectivity. *Light: Sci. Appl.* **2019**, *8* (1), 107.